\documentclass[conference]{IEEEtran}
\IEEEoverridecommandlockouts
\usepackage{cite}
\usepackage{amsmath,amssymb,amsfonts}
\usepackage{algorithmic}
\usepackage{graphicx}
\usepackage{textcomp}
\usepackage{xcolor}
\def\BibTeX{{\rm B\kern-.05em{\sc i\kern-.025em b}\kern-.08em
    T\kern-.1667em\lower.7ex\hbox{E}\kern-.125emX}}

\begin{document}

\title{SmartPubSub: Content-based Pub-Sub on IPFS}

\author{\IEEEauthorblockN{Pedro Agostinho}
\IEEEauthorblockA{
\textit{INESC-ID / Técnico Lisboa - ULisboa, Portugal}
}
\and
\IEEEauthorblockN{David Dias}
\IEEEauthorblockA{
\textit{Protocol Labs}}
\and
\IEEEauthorblockN{Luís Veiga}
\IEEEauthorblockA{
\textit{INESC-ID / Técnico Lisboa - ULisboa, Portugal}
}
}

\maketitle

\begin{abstract}
The InterPlanetary File System (IPFS) is a hypermedia distribution protocol enabling the creation of completely distributed applications. One of the most efficient and effective ways to distribute information is through notifications, with  a producer of content (publisher) sharing content with other interested parts (subscribers). IPFS already implements topic-based publish-subscribe systems under an experimental flag. The goal of this work is to advance on that, by developing a content-based pub-sub system (with subscriptions as predicates about event content) to disseminate information on top of IPFS in an efficient and decentralized way, leveraging its  infrastructure. We design two protocols: ScoutSubs  that is completely decentralized; FastDelivery that is centered in the publisher. With these two approaches,  we show the different advantages of having each of these protocols simultaneously by comparing ScoutSubs' full decentralization, and FastDelivery's centralization at data sources.
\end{abstract}


\section{Introduction}


Today's Web is managed by a few big players (Google, Amazon, Microsoft) that, throughout the years, added their mark to the Web and started incorporating smaller adversary companies to maintain their relevance. These major tech corporate giants make the current Web format highly centralized in a few data-centers own by them, making information easily subject to censure and allowing the use and control of their users' private data. 

The InterPlanetary File System~\cite{ipfs-filecoin-2020} was born to help create a decentralized World Wide Web, where users play a central role in the distribution and storage of information without the direct intervention of big corporate organizations. We can draw  parallels between what IPFS is trying to achieve and what web broadcast services like YouTube and others have done to diversify the  media, taking the market's control from the big media companies and placing it on the content producers.

IPFS content-addressing system is highly scalable and works with Kademlia's DHT and the Bitswap protocol~\cite{ipfs-bitswap-2021}. However, this mechanism is a static one meaning it always needs a search effort to trigger a data object retrieval. Another important detail about this static content-addressing is that the content is organized based on its physical properties, i.e. its raw bits and data type, that contribute to the content identifier (CID) of each data block (256 KB of maximum size). This is rather inflexible to adapt to high-popularity content and to improve locality and geo-distribution, and to reduce latency.

Currently, IPFS is still essentially a file-sharing system operating over a peer-to-peer network. Still, it has under an experimental flag a topic-based publish-subscribe system. Furthermore, in the IPFS community, some topic-based systems have also been created, e.g., Pulsarcast~\cite{ja-ipfs-pubsub-2021} and GossipSub~\cite{gossipsub} (that is the main one being employed). The nonexistence of a content-based publish-subscribe in IPFS and the Web is a significant lacking. Thus,  it opens space for our work to offer a content-based publish-subscribe alternative  on top of IPFS to enhance content selection, delivery, and  access performance.



\begin{figure}
    \centering
    \includegraphics[scale=0.75]{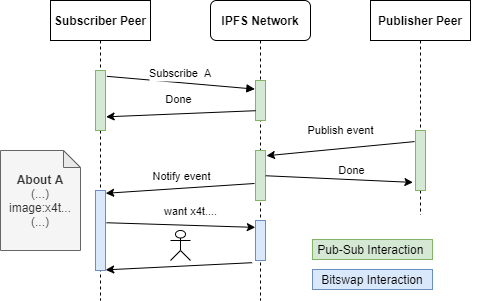}
    \caption{Peer interaction with  current static IPFS  and new dynamic publish-subscribe layer}
    \label{fig:staticdyn}
\end{figure}

In this work develop a scalable, flexible and performant  content-based publish-subscribe middleware, SmartSubs, to operate over IPFS. 
In particular, we provide a dynamic semantic-addressing layer (meaning content-addressing a human can understand) where the information is routed through the network depending on the users' interests, hence improving locality and latency over regular IPFS content retrieval. The publish-subscribe system is what allows such dynamic addressing, meaning  users express to the network their interests, and information of interest to them upon production is forwarded towards them. We aim at these two layers to work together, as in the example illustrated by Figure~\ref{fig:staticdyn}.

Previously, publish-subscribe and peer-to-peer have been addressed in common extensively (e.g., ~\cite{topic-pub-sub14},~\cite{arno-overlay-topic-pub-sub21}), but seldom on such widely deployed web-scale and web-based systems such as IPFS (an ecosystem that we aim to contribute to). In IPFS specifically, publish-subscribe has only been addressed  with a simpler, less expressive and flexible topic-based semantics~\cite{ja-ipfs-pubsub-2021,gossipsub}, as opposed to the content-based we are proposing. 

The rest of the document is organized as follows.  Section ~\ref{sec:arch}
describes our proposed solution. 
Section~\ref{sec:eval} describes the testbed platform used,  the 
relevant metrics and the results obtained. 
In Section~\ref{sec:rw}, we review previous work and systems in relevant related topics.
Some concluding remarks and extension proposals are presented in Section~\ref{sec:conc}.

\section{Architecture}
\label{sec:arch}

Our architecture was designed to be integrated with the existing IPFS routing overlay (Kademlia DHT~\cite{kademlia}, with the modifications proposed by S/Kademlia~\cite{Baumgart2007}). But there is a particularity with Kademlia peer ID distribution. In a network with Kademlia as its routing overlay, peers are assigned an ID and establish connections with others based only on ID distancing between them. That makes IPFS peer connection not geographically oriented, increasing the network's resilience, but reducing performance by allowing long redundant hops like US-CHINA-US-CHINA.
This results from the fact that Kademlia is a structured approach. 

To extend IPFS in order to support content-based publish-subscribe there is a spectrum of options between two main ones: i) extend, leverage it in some way, or build another layer structured approach on top of it, or ii) using it only as a bootstrapping mechanism, and then implement a fully non-structured (e.g. gossip-based) approach for the publish-subscribe.  We decided for an hybrid approach, i.e.: i) go \emph{structured},  by taking advantage of the efficiency of the existing structured approach in locating content, and leveraging the propagation paths of messages/content along the overlay; and ii) go \emph{unstructured} by trying to place popular content for premium users in a geographically oriented way, thus improving locality and performance (latency).
 
Our SmartPubSub is comprised of two subsystems that include each one its specific protocol and cater for different levels of service:
\begin{itemize}
\item \textbf{Decentralized Protocol - ScoutSubs:} This protocol  provides a content-based publish-subscribe system using the existing IPFS routing tables and adding a filtering structure inspired by the Hermes system~\cite{hermes} to disseminate events from publisher to subscribers based on their content and the content of the subscribers' subscriptions. ScoutSubs allows its users to possess a global semantic-based knowledge of their network.

\item \textbf{Publisher Centered Protocol - FastDelivery:} This protocol's focus is to deliver events as fast as possible. We provide an application-level multicast, geographically oriented, to provide low latency event delivery. The publisher coordinates its publish-subscribe service and may request its subscribers to disseminate its events making the IPFS' overlay used only for advertising. This protocol becomes interesting when subscribers are (more) interested in a (specific)  source of information instead (than) of a global notion of event content properties/predicates.
\end{itemize}

\begin{figure}
\centering
\includegraphics[scale=0.45]{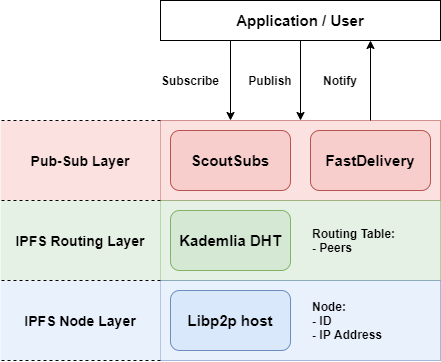}
\caption{Publish-subscribe Protocol Stack}
\label{fig:pubsubsuite}
\end{figure}

Our system's network stack is comprised of  three main layers illustrated in Figure~\ref{fig:pubsubsuite}. Briefly, at the bottom we have the libp2p host, which represents a node at IPFS and includes all its properties and information. For our publish-subscribe middleware, we will use a node's ID and its address.
Representing IPFS content routing we have an instance of its Kademlia DHT, in which we will use its routing tables to be able to reach the closest peer of every key (acting as a rendezvous node) in the network in a logarithmic number of steps.
Our publish-subscribe middleware layer will have both our protocols that will use the  ones below to provide users  and  external applications the possibility of publishing or subscribing to events in a content/meaningful way. 

\subsection{\textbf{Expressing Content}}

In both protocols, the way a subscriber expresses its (content-based) interests and the publisher expresses the content of the events it produces, is by assigning it a predicate. A \emph{predicate} is an expression assigned to a piece of data (event/message/file) that awards a semantic meaning to it, thus allowing more flexible and richer descriptions and selection of its content (as opposed to assigning it exclusively a tag, category or topic). This expression is composed by attributes that add specific meaning to a predicate.

In SmartPubSub, we combine  two types of attributes to characterize interest in content, i.e., to define a predicate:

\begin{itemize}
    \item \textbf{Topic:} should be single word key phrases capable of capturing the essence of the described data. Common examples can be names of countries, companies, bands, clubs or sports.
    \item \textbf{Range Queries:} are attributes with numerical meaning, composed by a characteristic, feature, property and its numerical interval/value. Common examples of these numerical characteristics can be price, temperature, height or dates. For simplicity, we assume there is prior agreement on the units used for each attribute in predicates (e.g. use dollars for prices and Celsius for temperature).
\end{itemize}

Thus, predicates need to be assigned to events published on the system and to the subscriptions, so that our publish-subscribe may understand who is interested in what. To simplify, predicates are assigned to events by their publishers and to subscriptions by the subscribers.

\subsection{\textbf{ScoutSubs Protocol}}

As mentioned earlier, the ScoutSubs protocol provides a publish-subscribe middleware over IPFS' content routing overlay. The base design of this system was inspired by Hermes~\cite{hermes}, for it shares a topic and a (predicate) filtering layer over it that provides a content-based approach.

\begin{figure}
\centering
\includegraphics[scale=0.45]{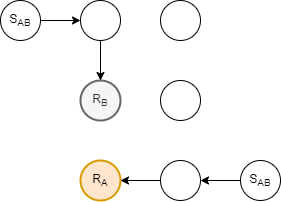}
\caption{Subscription Forwarding}
\label{fig:subgo}
\end{figure}

\paragraph*{\textbf{Rendezvous}} The topic-based layer is created by the rendezvous nodes. There are as many rendezvous nodes as attributes, being the one representing the attribute (e.g. \texttt{football}) the closest peer to the key generated by its string (usually hashing).
The purpose of these nodes is to provide a point of reference for the subscription forwarding. So, if we have the subscription \texttt{[football, Tom Brady]}, we first see which of these attributes has the closest ID to ours, and then we forward the subscription towards it, minimizing the number of hops. On the other hand, the publisher needs to send its events towards all the rendezvous nodes of its event attributes to ensure coverage. We limit the  effort to all (possibly hundreds or thousands)  subscribers, at the expense of extra effort from publishers for more complex events.Figure~\ref{fig:subgo} illustrates that.

Additionally, the path between the subscribers and rendezvous nodes needs to be backed up. All other functions and properties are inherited directly from Kademlia, i.e. the pathway from the publisher to the rendezvous node only uses information from the Kademlia's module, not needing any backup mechanism besides the fault tolerance already present at Kademlia's operations.

\begin{figure}
\centering
\includegraphics[scale=0.45]{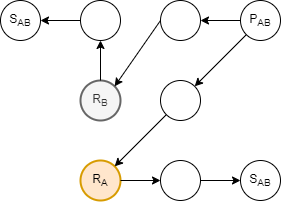}
\caption{Event Forwarding}
\label{fig:eventgo}
\end{figure}

\paragraph*{\textbf{Filtering}} To provide a content-based approach, we implemented a filtering mechanism over the rendezvous nodes. This way, before a subscription is sent towards the rendezvous node, it leaves its filter (subscription) at each intermediate node (i.e. leveraging the existing routes in the structured overlay that converge towards the rendezvous). Thus, when a publisher forwards the event to the rendezvous node, once it arrives at it, it will follow the reverse path of the subscriptions back to all the interested subscribers, as Figure~\ref{fig:eventgo} shows, minimizing message replication.

All the filtering information is kept at a filtertable which initially is simply a replica of the Kademlia routing table~\cite{kademlia}. Upon receiving subscriptions from those nodes, it adds filters to those node's entries. When receiving a subscription from a new node, it needs to create a new entry at the table and register the filter. Filters upon received can be also merged or ignored, when the result is the same in the forwarding process, to bound the size of the filtertables.

\subsubsection{\textbf{Event Forwarding}}
\label{sec:arch:scout:redirect}

\begin{figure}
\centering
\includegraphics[scale=0.38]{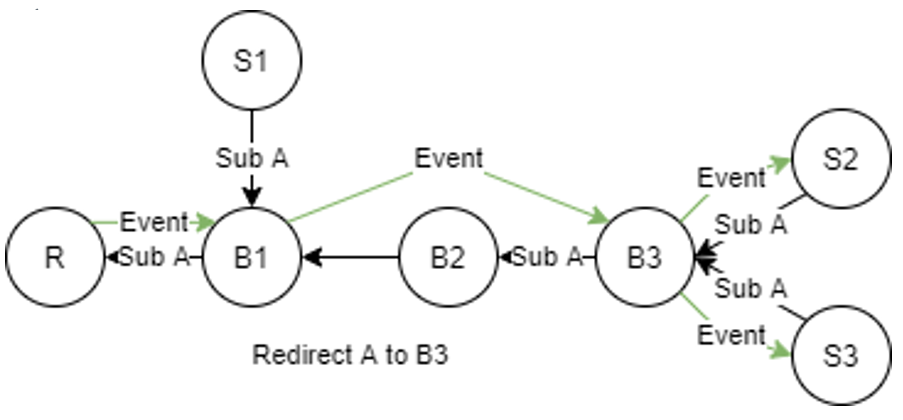}
\caption{Forwarding using shortcut}
\label{fig:redirect}
\end{figure}

To optimize event forwarding, we  add a mechanism that may allow an event to jump as many hops as possible (as a  shortcut) without compromising its delivery to all interested subscribers. To achieve this, when a node receives a filter from one peer towards a rendezvous node, it  always forwards upstream the option to provide a redirect (jump over itself), if there were no filters forwarded from other peers to that same rendezvous, as shown in Figure~\ref{fig:redirect}. We need then to keep track of how many filters were forwarded to each rendezvous. If the number of filters is below two, we may provide a shortcut option, but if the number gets equal to or larger than two, we must warn the node upstream that the shortcut is no longer valid.

The first immediate advantage is that we reduce the number of hops on the network, saving bandwidth and reducing event delivery time. The other important point is that the peers that are jumped over no longer have to check their filtertables, saving their precious CPU cycles. This fact is even more interesting because shortcuts are used more frequently on unpopular events. This means that searching a filter table with lots of filters to then have only one hit avoids wasting CPU unnecessarily.

\subsubsection{\textbf{Fault Tolerance}}
\label{sec:arch:scout:fault}

To tolerate the failure of any nodes along the dissemination path (rendezvous to subscribers), the filtering information must be backed up somehow. Thus, the information relevant to each node are the filters it contains in its filtertable, being then essential to back up the filtertable to those peer's backups. 
To achieve this, we decided that every node will have \emph{f} backups (a parameter with default value of 2 as our \texttt{faulttolerancefactor}) and that these are the peers closer to the node in question by ID (a typical approach in peer-to-peer replication). This backup mechanism allows SmartPubSub to tolerate \emph{f} failures of consecutive nodes by ID, meaning that in a large network where node failures occur independently of a peer's ID, SmartPubSub can withstand multiple and scattered node failures, provided that no more than \emph{f} nodes with consecutive IDs fail.
Backups of intermediate nodes do not need to be long-lived nodes because these are referenced directly and can  be refreshed upon change, when their failure is detected.

Additionally, the filtering information arriving at the rendezvous node also needs to be backed up, and it is so on the closest nodes to the rendezvous attribute(s) key(s) (instead of the closest ones to its ID), in order to ensure that the event information is backed up and reachable through any of its attributes. Another relevant  detail is that the node upstream (closer to the rendezvous) also needs to know the backup nodes of the peers it receives the filters from. Having this requirement means that each entry of the filtertable, besides having ID, endpoint address, and filters, also needs to have that node's backups endpoints.

\begin{figure}
\centering
\includegraphics[scale=0.38]{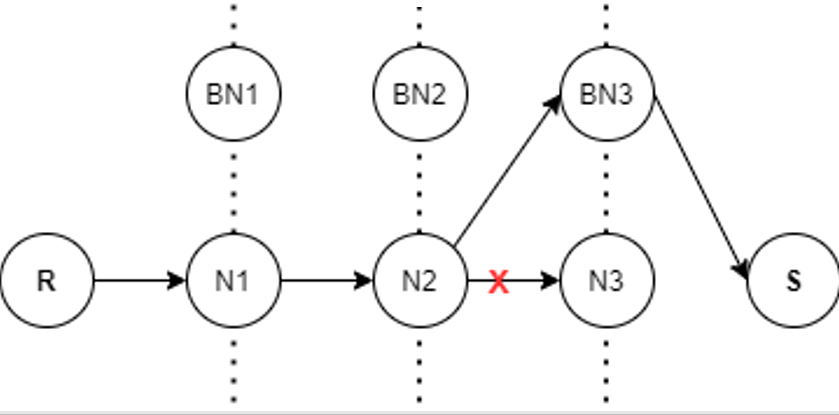}
\caption{Forwarding event using backup}
\label{fig:tolerance}
\end{figure}

With all of this, once an event arrives at a node, it first checks its filtertable to know which ones are interested in a particular event and then tries to forward it to them. If one node is not responding, it will send the event to the first working backup of that node. Once the backup receives the event with a backup flag activated, it will check the copied filtertable of the failed node and forward the event downstream. We can see the backup chain working in Figure~\ref{fig:tolerance}.

\subsubsection{\textbf{Ensuring Event Delivery}}
\label{sec:arch:scout:reliable}

When a peer crashes mid forwarding process, the subscribers downstream may not receive the event. To ensure this does not happen, we  implement a tracking mechanism resembling an acknowledgement chain. The challenging part is implementing this in a completely decentralized network without becoming inefficient.
Thus, between the publisher and the rendezvous node, the publisher will keep forwarding the event towards the rendezvous node until the latter receives it and sends an acknowledgment back. Before acknowledging the reception and process of the event to the publisher, the rendezvous node will track the event and send tracking requests to all its backups. Tracking means that the rendezvous will check its filtertable and create a map with all interested peers of that event. This way, once it receives an acknowledgment from every peer, the event will be considered as (eventually) successfully delivered, since  in the worst-case, it will only need to resend the event to the peers that have not confirmed yet.

Incrementally, and in decentralized manner, at all the regular (non-rendezvous) peers downstream, part of this mechanism is repeated: they keep a map with the interested peers, and upon receiving all acknowledgments, they forward their acks upstream. This approach also allows re-sending each event only to those peers who have not received it yet. The intermediate nodes have a passive role, being the rendezvous node, the manager of the re-sending process, the one that forwards the events back if their were not confirmed at all peers before a timeout.

\subsubsection{\textbf{Protocol Maintenance}}
\label{sec:arch:scout:maintenance}

To maintaining the protocol working over time, there needs to be  management of the filtering information. ScoutSubs does not allow explicit unsubscribing operations due to a subscription being a filter that can be merged or omitted. Therefore, we need to perform garbage collection of subscriptions no longer relevant by executing a refreshing routine.

To ensure that all subscriptions on the system are still relevant, we force the subscriber to resubscribe to the predicates it is still interested in every period of time \emph{t}. Every \emph{2t} time, each peer on the system will replace its main filtertable with a secondary one that was being compiled. A new secondary filter table is then created to receive new and resubscribed filters. Besides providing an option to abandon a previous subscription and avoiding unbounded growth use of storage usage, it also allows the system to regenerate completely its fault tolerance capacity back to its \emph{f} maximum every \emph{2t} time. 

\paragraph*{\textbf{Rendezvous maintenance}} The role of the rendezvous node has one particularity: besides being properly functioning, it also needs to be a long-lived node, meaning it needs to be working for an entire refreshing cycle. This run-time requirement is necessary since a new node entering the network, and closer to the key than its previous rendezvous node, will become the new rendezvous node for that attribute. But the new rendezvous node will be missing most of the filtering data; so, until it has  completed one refreshing cycle, besides sending to the ones it already knows, it needs to redirect the events to an old rendezvous or one of its old backups.

\subsection{\textbf{FastDelivery Protocol}}
\label{sec:arch:fast}

As previously mentioned, FastDelivery's main objective is to disseminate events as fast as possible. To do so, we need to go beyond IPFS' overlay structure and, in part, centralize the event dissemination at the publisher. The publisher can still provide events to the subscribers via ScoutSubs but must manage a group of premium subscribers, to which it sends events directly, or in at most two geographically oriented hops (overlay hops). Premium subscribers need to provide the publisher their endpoint, location (Region/Country), and resources (network and CPU-wise). 

\paragraph*{\textbf{Motivation}} The reason of designing this protocol alongside ScoutSubs was to reflect when a more centralized approach to disseminate information is actually the best option. In this case, when a subscriber is not searching just for a topic/content of a publication but a particular publisher with a certain reputation or popularity, this alternative becomes an option offering better performance,  i.e. lower latency. As our main goal is to design a decentralized content-based publish-subscribe over P2P, we decided to develop a protocol as simple as possible just to assess the feasibility of the option above.

\paragraph*{\textbf{Overall Design}}
\label{sec:arch:fast:design}

In this protocol, a publisher manages multicast groups: a data structure containing its interested subscribers and their subscriptions predicates. Each multicast group is represented by its publisher ID and the group's predicate (apple/france/price[0,1]).

To manage the multicast group's subscribers and recruit them if the publisher needs assistance, we decided to group subscribers into regions, ordered by capacity. Thus, once one publisher gets too many subscribers, the most powerful one gets recruited as a helper to assist the publisher.

To organize the subscribers' predicates, we save their subscriptions in a simple list, in the case of all predicate's attributes being of the topic type. If there are any range type attributes, we use a binary tree to organize the subscription. If the multicast group's predicate has several range attributes, it will need the same number of range trees and go through their query results. 

For simplicity, we consider a helper's capacity as the number of other subscribers it can help the publisher manage. After agreeing to assist the publisher, the helper will only support the structure that manages the subscriptions' predicates to forward the publisher's events to the interested subscribers. This support structure is a list or range trees with the subscribers delegated by the publisher. 

In terms of advertising a publisher's multicast group, we use advertisement boards at the rendezvous nodes of the attributes of the group's predicate, to ensure it can be discovered by any subscriber. But a publisher may also prefer to keep its endpoint address private.

\section{Evaluation}
\label{sec:eval}

We implemented SmartPubSub in golang, and developed several variants with and without the tracking and redirect mechanisms \cite{respage}. We built a testing environment using testground \cite{testground} and set up several several testing scenarios to test our pubsub.
 The following results were achieved using an Ubuntu VM with 126GB of RAM and a 16-core CPU. Besides the graphics and data presented here, more detailed data of this and other experiences can be found on our results page \cite{respage}.

\subsection{\textbf{Variant's testing}}
\label{sec:eval:res:var}

Here we present the results of each variant tested through different scenarios. The goal is to analyze if the redirect and reliable mechanisms are working and are not inefficient. This test battery allowed us to analyze our system  in the development phase, and to correct some bugs it helped to detect. Each test run in this section had a 60 node network, where each node is running inside a Docker container, executing a complete host with the full IPFS and SmartPubSub stack (i.e. there is no peer-to-peer simulation taking place). 

\paragraph*{\textbf{Correctness}} To know how correct our publish-subscribe is, we compared the duplicated and missing events by each variant at each scenario. In all variants and scenarios, our publish-subscribe performed with 100\% reliability and produced in all variants some duplicated events. The FastDelivery approach had perfect results for it manages its subscribers directly, resource consumption was slightly lower than the other variants. Because our experience used a 60 node network, and each node had around 30-40 peers, communication is mostly direct (2-4 hops) and redundant paths are common.

\paragraph*{\textbf{Event Latency}}The results from a normal scenario were an average event latency for the used network composition and configuration of 200-250 ms. In a scenario with subscription being made at the time of publishing, the results of the average event latency were of 200-250 ms. In a scenario with 10 times more events being published than normally, the average event latency was of 1780-2500 ms. In a scenario with 2 failing peers, the average event latency was of 200-270 ms. These results reflect that the higher the event production is, the higher the event latency will be.

\paragraph*{\textbf{Resource Consumption}} Looking at the memory and CPU consumption of our publish-subscribe during our test runs, an higher memory usage is followed by a higher CPU time usage. Each scenario has different periods, as in the event burst scenario, where its testing period is around 12 seconds, and the normal one is less than half that. 

When looking to  Table~\ref{table:5}, we can see that the reliable variants, especially in the event burst scenario, have a substantially higher CPU and memory usage. The reasons for this are the extra relation between tracker and rendezvous node and the management of the acknowledge chains. The impact of the redirect mechanism is not palpable with a smaller network, and so we cannot comment on its performance.

\begin{table}
\small
\renewcommand{\arraystretch}{1.5}
\centering
\begin{tabular}{ |p{1.5cm}||p{1cm}|p{1cm}|p{1cm}|p{1cm}|}
\hline
Variant & Normal & Sub Burst & Event Burst & Fault \\
\hline
Base-Unreliable & 57.3 MB 4.01 s & 36.0 MB 2.36 s & 173 MB 11.45 s & 19.5 MB 1.94 s \\
\hline
Redirect-Unreliable & 116.9 MB 8.13 s & 26.7 MB 3.82 s & 179 MB 14.62 s & 5.10 MB 2.15 s \\
\hline
Base-Reliable & 99.2 MB 6.54 s & 62.0 MB 3.88 s & 310 MB 20.21 s & 22.5 MB 3.46 s \\
\hline
Redirect-Reliable & 108.3 MB 8.31 s & 37.0 MB 3.61 s & 268 MB 21.38 s & 21.5 MB 3.84 s \\ 
\hline
\end{tabular}
\caption{Average Memory and \textbf{CPU} user-time used per node}
\label{table:5}
\end{table}

\subsection{\textbf{Replication Performance}}
\label{sec:eval:res:replica}

After analyzing each variant of our publish-subscribe, we decided to test with our Redirect-Reliable variant how its performance changes if we increase the \texttt{faulttolerancefactor}. We also took the chance to analyze how the system performs with an increase in the number of subscriptions each subscriber does. We vary the publish-subscribe's replication from 1, 2, ,3 and 5 in a 75 node network, set up and running as described earlier.

\begin{figure}
\centering
\includegraphics[scale=0.40]{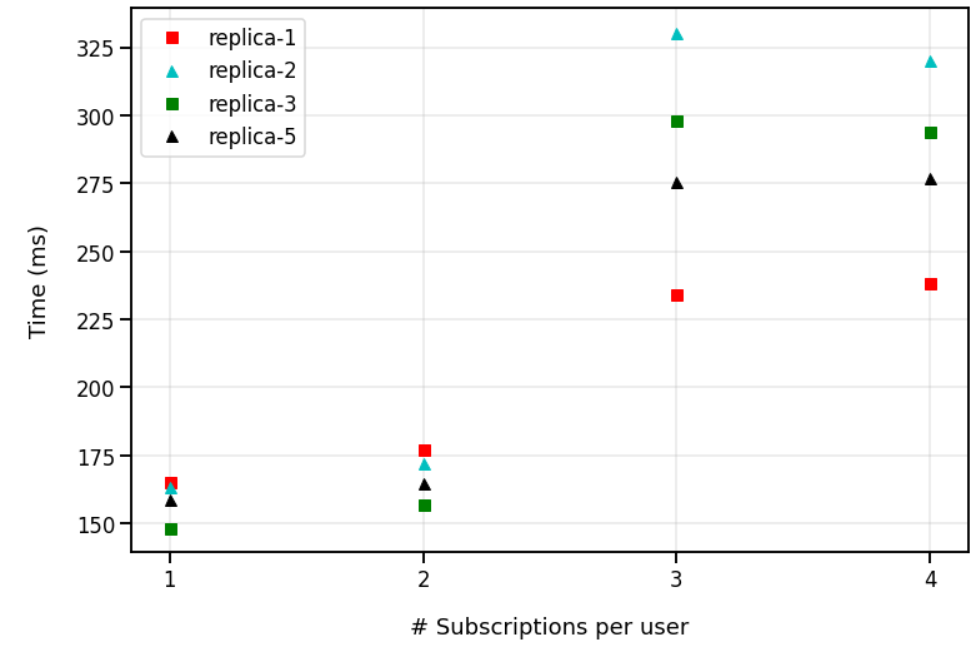}
\caption{Average Event Latency per replica factor at each stage}
\label{fig:lateventfinal}
\end{figure}

\begin{figure}
\centering
\includegraphics[scale=0.40]{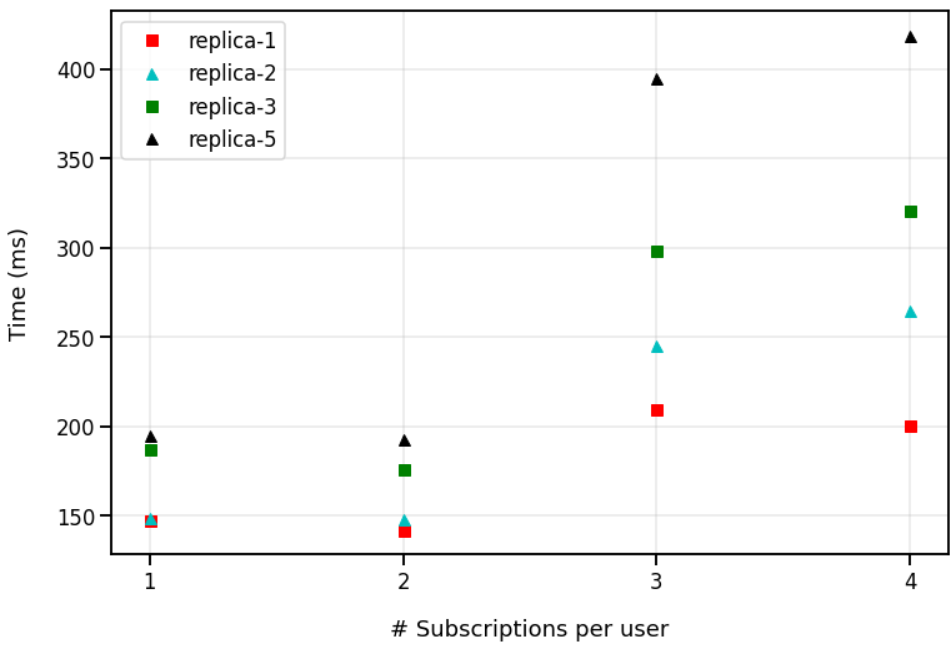}
\caption{Average Subscription Latency per replica factor at each stage}
\label{fig:latsubfinal}
\end{figure}

\paragraph*{\textbf{Event and Subscription latency}} We can start by looking at  Figures~\ref{fig:lateventfinal} and~\ref{fig:latsubfinal} to analyze the results regarding the event and subscription latency, respectively.

The results of our publish-subscribe subscription latency are as predicted since the bigger the replication factor, the longer it takes to complete a subscription. The same can be said of the number of subscriptions per subscriber, since each subscribing operation needs to check all the filters of a filtertable entry. Filter checking is necessary to enable merging  a subscription filter with others, or ignore it (because one of those in the entry already encompasses it) in order to bound the storage requirements of subscription filtering at each node.

When looking at the event latency, we see that the correlation is not as strong as in subscriptions. A subscription needs to be sent to a node's backups, and they need to add and summarize the subscription filters. In an event forwarding, the only interaction between the main path and the backups is at the rendezvous between the different trackers. 

The independence between fault-tolerance and event forwarding mechanisms in a non-failure scenario was designed explicitly to achieve  faster event forwarding (much more frequent) to the detriment of the subscription operation (less frequent). Event latency is dependent on the order of the filters in the at each filtertable entry, because if the first filter matches the events, that node will not have to check the other ones (this leaves room to further optimization of filtertables to favour earlier match of events or early discard of events, taking into account their coverage).

\begin{figure}
\centering
\includegraphics[scale=0.40]{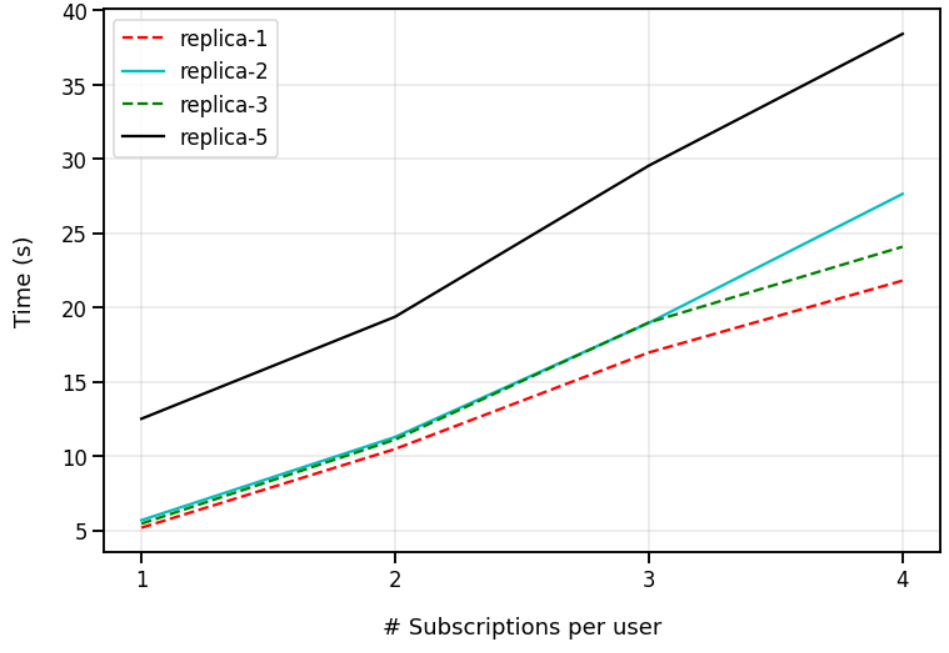}
\caption{Accumulated CPU used for each replica factor}
\label{fig:cpufinal}
\end{figure}

\begin{figure}
\centering
\includegraphics[scale=0.40]{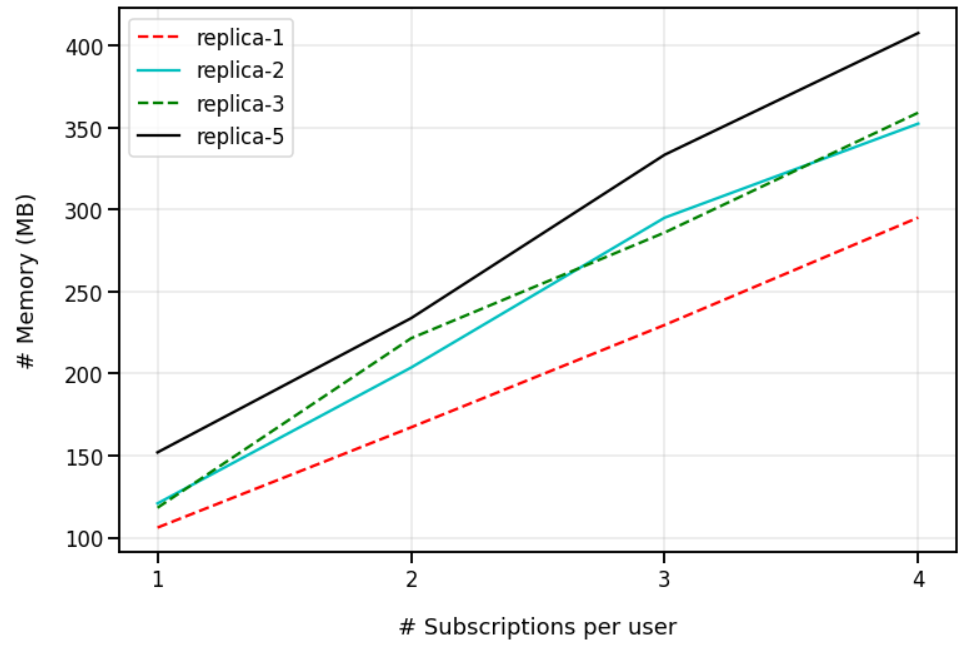}
\caption{Accumulated Memory used for each replica factor}
\label{fig:memoryfinal}
\end{figure}

\paragraph*{\textbf{{CPU} and memory usage}} We now analyze the resources used in this experience by checking Figures~\ref{fig:cpufinal} and~\ref{fig:memoryfinal} for the {CPU} and memory usage, respectively.
{CPU} usage does not increase with the number of subscriptions per user but tends to increase with the replication factor.
Total Memory consumption is straightforward since it increases with the replication factor and remains mostly constant with the increase of the number of subscriptions since their size is small. 

\paragraph*{\textbf{Scalability}}

Regarding the scalability of our system, when analyzing our system's performance with the increase of subscriptions per user, we can confirm that in terms of resource usage, the system is scalable. When looking at the memory and {CPU} usage, we can see that they do not follow the increase in subscriptions as the graphic tends to a linear regression. 
Being an accumulative graphic, it means that resource consumption is approximately the same at each stage. 

In terms of performance, we can see a change, although far from affecting the user perception of the event speed delivery. This may be further  reduced when running in a larger network with more spaced rendezvous and distributed workload.





\section{Related Work}
\label{sec:rw}

In this section we address literature relevant to our work. We start with  the different characteristics of \textbf{Peer-to-Peer} systems, in particular content distribution ones such as IPFS. Afterwards, we address  the role and characteristics of \textbf{Publish-Subscribe} systems

\subsection{\textbf{Peer-to-Peer Content Distribution}}

A P2P system is commonly seen as one for decentralized sharing of computational resources over the network, maintaining  proper functioning even in the presence of  node failures, connectivity problems, and churn.
P2P networks are highly scalable and fault-tolerant because each node can act  both  as server and client, and so, the number of servers grows linearly with the number of clients, preventing  bottlenecks.

There are several P2P application types, but we are mainly interested in \textbf{content distribution systems} (the category that encompasses IPFS). These are designed to share  digital media and other data among users, ranging from simple direct file-sharing applications to sophisticated systems that create a distributed storage medium to securely and efficiently publish, index, search, retrieve and update  data.

Such systems have been  characterized~\cite{p2p-acm-2004} along two  axis, regarding their architecture: i) \textbf{structure}, and ii) \textbf{level of decentralization}. \textbf{Structure} pertains to how the overlay is organized. In structured approaches, content is placed in specific locations of the overlay ID space, making data lookup faster and more efficient, at the expense of forfeiting locality regarding geography. In unstructured approaches, the placement of content is unrelated to the overlay's topology, with data look-up potentially slower, but better tolerating environments with transient node population, more easily leveraging locality.

 The \textbf{level of decentralization} can be higher or lower across:

\begin{itemize}
    \item \textit{i) Purely Decentralized:} all nodes in the network perform the same tasks, acting both as servers and clients, and there is no central coordination of activities. Some  examples are Kademlia~\cite{kademlia} (IPFS overlay), and CAN \cite{can};

    \item \textit{ii) Partially Centralized:} is similar to purely decentralized systems, although some nodes assume a more important role, e.g. acting as local indexes for files shared by neighboring peers. Since these \textit{super-nodes} can be dynamically (re-)assigned by the overlay, they are not single points of failure. One example of this is Kazaa~\cite{kazaa};

    \item \textit{iii) Hybrid Decentralized:} in these systems, there is indeed a central server facilitating  interaction between peers, e.g. by maintaining directories of metadata, describing  files stored by  peer nodes, performing look-up. Still, the end-to-end interactions take place directly between two peer nodes. One such example  is Publius~\cite{publius}.
\end{itemize}

IPFS inherits the advantages of Kademlia by virtue of being purely decentralized (and thus highly scalable) and structured (therefore with efficient look-up). But users can only access content in IPFS whose specific CID they already know (or get through a name system) which is restrictive and inflexible. Existing topic-based publish-subscribe offers on IPFS~\cite{ja-ipfs-pubsub-2021,gossipsub} only partially address such limitations and inefficiency (i.e. inability to define interest in content in fine-grained manner).

\subsection{\textbf{Publish-Subscribe}}

Publish-Subscribe~\cite{manyfaces} is a message paradigm that provides complete decoupling between data consumers and data producers both in time, space and synchronization. The  main actors are the publisher (producer of events), the subscriber (consumer of specific events), and the information bus which needs to be distributed in large scale settings. This last one is the medium where subscriptions made, and published events, are forwarded;  the medium finds a way to notify the interested subscribers of the published events.
There are three types of publish-subscribe systems regarding the granularity of the subscriptions and events identifiers and/or predicates:

\begin{itemize}
\item \textit{Topic-based}: Here,  events and subscriptions are addressed with a topic. This approach is the simplest and easiest to implement, with  a topic being a keyword identifying the event or subscriber interest. Examples of topic-based publish-subscribe architectures include  Rappel~\cite{rappel}, SDN-like pub-subs~\cite{topic-bucket-pubsub-2020}, MinAVG-k-TCO~\cite{arno-overlay-topic-pub-sub21}, POSSUM~\cite{possum-2022},HoP-and-Pull (HoPP) for IoT environments~\cite{mob-pubsub-cn-2022};

\item \textit{Content-based}: Sometimes a topic does not fully/accurately describe the interest of a subscriber, resulting in receiving several unwanted events. A content-based approach will prove more precise and often more efficient w.r.t. bandwidth consumption. Here, subscriptions are represented by predicates that not only represent topics and subtopics but also range queries. Examples of content-based publish-subscribe  are Hermes~\cite{hermes}, Wormhole~\cite{wormhole}, PopSub~\cite{popsub-pubsub-2017}, PSDG~\cite{delivery-pubsub-2020};

\item \textit{Type-based}: With a type-based approach, an event is strictly characterized by a schema, enabling  closer integration with the language and the middleware code. Moreover, type safety can be ensured at compile-time by parameterizing the resulting abstraction interface with the type of the corresponding events. One example of a type-based publish-subscribe system is FlexPath~\cite{flexpath}.
\end{itemize}

\paragraph*{\textbf{Architectural Details}} Several types of publish-subscribe systems have been designed and implemented. In large-scale ones, akin to P2P,  one main characteristic is the centralization degree: pub-subs may be implemented: i) using a central server, ii) distributed servers or meshes, or iii) with a completely decentralized approach where nodes have the same role. The more centralized the approach, the less efficient is the routing and the structure is less tolerant to failures; on the other hand, reliability and persistence of  events is more easily guaranteed and algorithms are lighter and  easily implemented.
  
The other main characteristic is the \textbf{forwarding strategy} employed by the publish-subscribe w.r.t. events and subscriptions, which can follow one  of these main alternatives:

\begin{itemize}
    \item \textit{Rendezvous:} in this approach, some nodes on the network will provide a point of contact between publishers and subscribers. Rendezvous can be assigned for each type of event by hashing an event header or content to a network's addressable space (e.g. in structured P2P) or by checking a predefined list of rendezvous nodes.

    \item \textit{Filtering:} this approach can be used in different scenarios, but its goal is to transform a subscription into a filter. This way, once a node receives a filter, it may attempt to merge it with previous ones, reducing the number of filters it needs to check. This approach needs to implement a mechanism of distribution of the filters, propagating the filters towards  rendezvous and/or event publishers.

    \item \textit{Gossip:} one alternative approach is to diffuse events via gossiping. Here subscribers are grouped by interest, and publishers somehow forward their events to some of the interested subscribers. These subscribers then guarantee the event's forwarding among the remaining interested ones, relying on the ability of the subscriber to find the best neighboring peers based on its interests.

    \item \textit{Flooding:} the last  one is flooding  events or subscriptions over the entire network, requiring a cache to prevent event duplicated delivery. This approach is the easiest to implement (e.g. in unstructured P2P), but it wastes bandwidth  forwarding unwanted events,   not scaling well.
\end{itemize}


SDN-like publish-subscribe~\cite{topic-bucket-pubsub-2020} creates specific overlays regarding each topic for event propagation, with SDN-based systems in mind.
HoP-and-Pull (HoPP) addresses topic-based publish-subscribe for IoT ecosystems~\cite{mob-pubsub-cn-2022}, where network decentralization is limited, employing  RIOT-OS and IoT LoWPAN.
Topic-connected overlays (TCO)~\cite{arno-overlay-topic-pub-sub21} introduces \textit{k-topic-connected} overlays to ensure efficient propagation topologies and reliable event propagation, as long as fewer than k nodes interested in the same topic fail simultaneously.
They are complementary to SmartPubSusb as they do not address the web ecosystem and are not content-based (neither  
Pulsarcast~\cite{ja-ipfs-pubsub-2021}, GossipSub~\cite{gossipsub} over IPFS). 
SmartPubSub also builds multi-topic-overlays around rendezvous (one per topic in subscriptions), employs replication, and  optimizes event forwarding by merging filters (content-based  predicates).

POSSUM~\cite{possum-2022} targets IoT, using triple-vectors (combining topics and queries in knowledge graphs with RDF) to optimize subscription filtering with clustering and normalization.
PSDG~\cite{delivery-pubsub-2020} introduces delivery guarantees, without hampering throughput, in content-based publish-subscribe systems with distributed broker/mesh architectures.
PopSub~\cite{popsub-pubsub-2017} takes into account event popularity and brokers' resources in event propagation to improve resource efficiency in brokers, and reduce delivery latency in high-load.
These systems are complementary to SmartPub. Although content-based,
 they are not designed to address decentralized web-scale environments,
such as  SmartPubSub (based on Kademlia/IPFS). They optimize subscription filtering but employ centralized coordination in IoT, or distributed/hierarchical broker topology (wholly known in advance), to achieve high-throughput and not geo-scale. 

%





\section{Conclusions}
\label{sec:conc}

We propose a novel publish-subscribe middleware, SmartPubSub,  to provide  global content/semantic-addressing  over IPFS. It offers  two protocols: ScoutSubs and FastDelivery.
%
%
%
%
%
%
%
%
With the first, we find that a decentralized content-based publish-subscribe solution for the large scale web is  relevant and lacking. We propose a  global system that is not only physically content-based oriented where information is stored (as in IPFS), but one that truly provides a semantic content-based approach, with information forwarded through the web depending on its semantic content and user interest.
But decentralization is not always perfect. We address it with a protocol showcasing where some centralization can be advantageous.



\vspace{6pt}\footnotesize{\textbf{Acknowledgements}
This work was supported by national funds through FCT, Fundação para a Ciência e a Tecnologia, under project UIDB/50021/2020.}

\medskip

\bibliographystyle{unsrt}
\scriptsize{\bibliography{ref}}
\clearpage

\end{document}